\documentclass[aps,prx,reprint,groupedaddress]{revtex4-1}
\usepackage{graphicx}
\usepackage{amssymb,amsmath}

\begin{document}

%\preprint{}

\title{Photonic Topological Insulating Phase Induced Solely by Gain and Loss}


\author{Kenta Takata}
%\email[]{takata.kenta@lab.ntt.co.jp}
%\thanks{}
\affiliation{NTT Nanophotonics Center, NTT Corporation, 3-1 Morinosato-Wakamiya, Atsugi 243-0198, Kanagawa, Japan}
\affiliation{NTT Basic Research Laboratories, NTT Corporation, 3-1 Morinosato-Wakamiya, Atsugi 243-0198, Kanagawa, Japan}

\author{Masaya Notomi}
%\email[]{}
%\thanks{}
\affiliation{NTT Nanophotonics Center, NTT Corporation, 3-1 Morinosato-Wakamiya, Atsugi 243-0198, Kanagawa, Japan}
\affiliation{NTT Basic Research Laboratories, NTT Corporation, 3-1 Morinosato-Wakamiya, Atsugi 243-0198, Kanagawa, Japan}


\date{\today}

\begin{abstract}
We reveal a one-dimensional topological insulating phase induced solely by gain and loss control in non-Hermitian optical lattices. The system comprises units of four uniformly coupled cavities, where successive two have loss, the others experience gain and they are balanced under two magnitudes. The gain and loss parts are effectively \textit{dimerized}, and a bulk bandgap, topological transition, midgap topological edge and interface states in finite systems can all be achieved by controlled pumping. We also clarify non-Hermitian topological numbers and edge states in gapless conditions.
\end{abstract}

%\keywords{}

\maketitle


%\section{Introduction}
Controlling optical properties with external signals is a major destination in photonics research \cite{Photonics}, and it is largely associated with tailoring the refractive index.
Recent studies have revealed that the imaginary part of the refractive index, namely gain and loss, can do much more than just tuning the optical intensity.
The key concept, parity-time ($\mathcal{PT}$) symmetry \cite{PTSym,PTSym2}, was introduced for obtaining real spectra of quantum systems with non-Hermitian components. 
Its analogy in optics \cite{PTOpt1, PTOpt2} corresponds to complex refractive index profiles with symmetric real parts and antisymmetric imaginary parts, i.e. $n({\bf r}) = n^{*}(-{\bf r})$.
Such a system can show an exceptional point (EP) \cite{PTEP} where its eigen-detuning sharply changes from real to imaginary values \cite{PTObs09,PTObs10} (spontaneous $\mathcal{PT}$ symmetry breaking).
There are many interesting phenomena related to $\mathcal{PT}$ symmetry including power oscillation \cite{PTOpt1,Breathers}, double refraction \cite{PTOpt1,PTFBLattice}, Bloch oscillation \cite{BlochT,BlochE}, mode-locking \cite{ModeL}, coherent absorption \cite{CPA,CPALaser,CPALaserExp}, fast light \cite{PTFBLattice,PTHoneycomb,PTBHCROW}, and unidirectional reflectivity \cite{PTUITheor,PTUIExp,USS}.
Furthermore, nonlinearity-induced isolation \cite{PTUT1,PTUT2}, single-mode lasing \cite{PTLaser1,PTLaser2}, and beam steering \cite{PTVCSEL} were achieved under controlled pumping.

To widen the scope of non-Hermitian optics \cite{NHOReview1,NHOReview2}, there are growing attempts to incorporate topological features to photonic systems with gain and loss. 
While Hermitian photonic topological phases \cite{TPhC,TPhoton1,TPhoton2} are based on celebrated discoveries such as the quantum Hall effect \cite{QHE,TKNN} and topological insulators \cite{Z2TO,TICQ}, non-Hermitian topological optics originates from the theoretical question as to whether or not stable topological quantum states exist in non-Hermitian systems \cite{TRSNH,NHedge,AbsTI,PTAATopo,DMFTopo}.
For photonics based on classical electromagnetic waves, however, it has been clarified that there exist topological states even when their eigenvalues are not real \cite{NHLTopo}.
Researchers applied Su-Schrieffer-Heeger (SSH) photonic lattices \cite{SSHModel} with relevant loss and experimentally confirmed their topological interface states \cite{NHLTopoExp} and topological transition \cite{BulkPhTT}. 
Moreover, a topological bound state with global $\mathcal{PT}$ symmetry was observed in a waveguide array \cite{BoundPhTS}.
Even the lasing of photonic topological edge states has recently been shown to be feasible \cite{1DTopoLaser1,1DTopoLaser2,1DTopoLaser3,2DTopoLaser}. 

Then, another question may arise. 
\textit{Can we create a topological insulating phase solely from gain and loss control?} 
In previous studies of non-Hermitian optics, the emergence of nontrivial topologies was attributed to Hermitian factors, namely the magneto-optic effect \cite{2DTopoLaser}, and lattice and coupling profiles of host systems \cite{PhSSH,AAModel,NHAES,ORATI,NHedge}.
Even though such systems are armed with non-Hermiticity, they take over original Hermitian topological characters predetermined by fabrication.
Moreover, gain and loss in conventional non-Hermitian systems \cite{CMTLaser,NHOReview1,NHOReview2} only close frequency bandgaps. Thus, they were considered to destroy topological insulating phases.
In contrast, our aim is to generate a topological bandgap solely by adding static gain and loss to a topologically trivial structure.
Then, achieved topological features, including a well-defined topological number, should originate purely from non-Hermitian factors.
Here, the gain and loss are readily tunable by injection current with independent electric channels or properly masked optical pumping in various laser systems \cite{PTBHCROW,PTVCSEL,1DTopoLaser1,1DTopoLaser2,1DTopoLaser3}.
We will hence have full manipulability over the topological properties in optical circuits, such as a topological transition, and the number and position of topological states, simply by changing the gain and loss. 
The resultant potential of high-frequency modulation of the photonic topology will open up a new horizon for topological engineering.

Here, we show theoretically a one-dimensional photonic lattice with the gain- and loss-induced reconfigurable topological insulating phase.
We consider unit cells of four uniformly coupled resonators, with loss introduced into two successive cavities and gain introduced into the other two.
The system then forms a pair of dimers by effective decoupling between cavities with gain and loss.
This can result in a bulk bandgap, topological transition and midgap edge states for a wide range of parameters.
Topological interface states can also be achieved at a controlled boundary between the nontrivial and trivial lattices.
Our scheme is unique in non-Hermiticity-based \textit{midgap} topological states protected by their isolation from bulk states.
Although, we also clarify system topological features in non-Hermitian gapless conditions, which will be relevant with defect and edge states in gapless systems \cite{TDSBulkTrivial1,TDSBulkTrivial2,2DAnomoulousTI,NHedge}.

\begin{figure}[htbp]
\includegraphics[width=8.6cm]{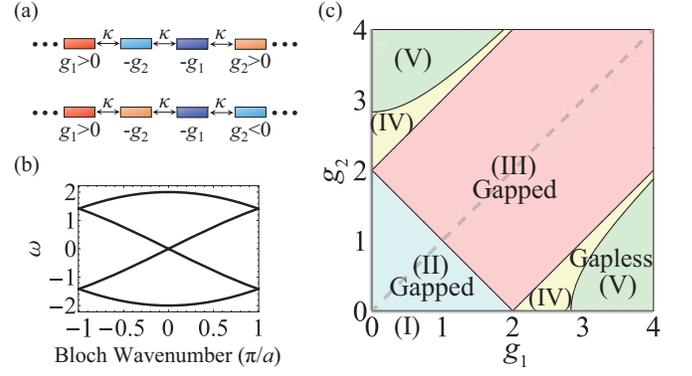}
\caption{(Color online) (a) Schematic of considered system. Upper and lower lattices for $g_2 > 0$ and $g_2 < 0$ are topologically nontrivial and trivial, respectively. (b) Folded cosinusoidal band structure for $g_1 = g_2 = 0$. (c) Phase diagram for the system band structure. $\kappa = 1$, $g_1, g_2 \ge 0$.}
\label{fig:schematic}
\end{figure}

\textit{Theoretical model.--}
The system comprises periods of four single-mode cavities with uniform couplings $\kappa$ [Fig. \ref{fig:schematic} (a)].
We introduce an on-site imaginary potential profile $(i g_1, - i g_2, -i g_1, i g_2)$ to the cavities, where its positive and negative coefficients mean gain and loss, respectively.
Here, we assume that $\kappa$, $g_1$ and $g_2$ are sufficiently small compared to the cavities' resonant frequency $\omega_0$ and the cavity-mode Q-factor is high, so that we can safely neglect the effect of the imaginary index profile and radiation loss on $\kappa$, as expected in semiconductor lasers \cite{PTUT1,PTUT2,PTLaser2,1DTopoLaser1,1DTopoLaser2,1DTopoLaser3,PTVCSEL,PTBHCROW,gainCROW}.
Within the linear analysis, the coupled mode equation describing the system is equivalent to the Schr\"{o}dinger equation $i \partial _t \left| \Psi \right\rangle = \hat{\mathcal{H}} \left| \Psi \right\rangle$, where $\left| \Psi \right\rangle = (\{ \Psi _n \} )^{\rm T}$ is the vector of the slowly-varying complex cavity-mode amplitudes ($n$: cavity index) and $\hat{\mathcal{H}}$ is a tight-binding lattice Hamiltonian.
Considering the Bloch theorem and a dynamical factor $e^{-i \omega t}$, the analysis reduces to an eigenvalue problem for the four-component eigenvector $\left| \psi _{\rm B} \right\rangle$ under the Bloch Hamiltonian $\hat{\mathcal{H}}(k)$,
\begin{equation}
\hat{\mathcal{H}}(k) = \left( \begin{array}{cccc}
i g_1 & \kappa & 0 & \kappa e^{-i k a} \\
 \kappa & -i g_2 & \kappa & 0 \\
0 & \kappa & -i g_1 & \kappa \\
\kappa e^{i k a} & 0 & \kappa & i g_2 \\
\end{array} \right), \label{eq:Hamiltonian}
\end{equation}
where $a$ is the spatial interval between the four-cavity units and $k$ is the Bloch wavenumber.
The eigenfrequency detuning $\omega (k)$ with reference to $\omega_0$ is given by,
\begin{equation}
\omega (k) = \pm \frac{1}{\sqrt{2}} \sqrt{A \pm \sqrt{ A^2 - B^2 - 16 \kappa^4 \sin^2 \frac{k a}{2}}}, \label{eq:eigenvalue}
\end{equation}
where $A = 4 \kappa^2 - g_1 ^2 - g_2 ^2$ and $B = 2 g_1 g_2$.
We also find analytic forms of $\left| \psi _{{\rm B},s} \right\rangle$ ($s$: eigenstate index), although they are too complicated to be given here.
In the following analysis, the gain and loss are measured with respect to the cavity coupling, i.e. $\kappa = 1$.
We focus on the case where $g_1 > 0$ and $g_1 \ge |g_2|$ for studying the bulk properties, because the spatial and/or time reversal can map the system with this condition to that with the other parameter range.
When $g_1 = g_2 = 0$, the system has a gapless four-fold cosinusoidal band structure with a degeneracy at $\omega(0) = 0$, because of the reduced first Brillouin zone [Fig. \ref{fig:schematic} (b)]. 

The system band structure is classified into five patterns via the value of the inside of the double radical sign of Eq. (\ref{eq:eigenvalue}) for $k = 0$ [Fig. \ref{fig:schematic} (c)].
With $g_1, g_2 \ge 0$ for simplicity, the divided phase regions are, (I) $B=0$, (II) $A+B>0$, $A-B \ge 0$, (III) $A+B>0$, $A-B<0$, (IV) $A+B \le 0$, $A^2 - B^2 - 16 \kappa^4 < 0$ and (V) $A^2 - B^2 - 16 \kappa^4 \ge 0$.
Here, the phase boundaries are symmetric to $g_1 = g_2$.
The systems in phase (II) and (III) have complete frequency bandgaps, while those in phase (I), (IV) and (V) are gapless.
System band structures for phase (IV) and (V) are shown in Supplementary Materials \cite{Supplementary} (Fig. S1).
Note that the diagram for $g_2 < 0$ is obtained by the mirror inversion of Fig. \ref{fig:schematic} (c) with regard to $g_2 = 0$.

In Eq. (\ref{eq:eigenvalue}), we see that $\omega (k)$ is real as long as $A>0$ and $A^2 - B^2 - 16 \kappa^4 \sin^2(k a/2) > 0$.
Such real eigenvalues are obtained because $\hat{\mathcal{H}}(k)$ has a pseudo-Hermiticity \cite{PHEigenReal}, $\hat{\mathcal{S}}(k) \hat{\mathcal{H}}(k) \hat{\mathcal{S}}(k) = \hat{\mathcal{H}}^{\dagger}(k)$.
Here, the $k$-dependent linear operator $\hat{\mathcal{S}}(k) = \hat{\mathcal{S}}(k)^{-1} = \hat{\sigma} _x \otimes (\cos \frac{k}{2}) \hat{I_2} + \hat{\sigma} _y \otimes (\sin \frac{k}{2}) \hat{I_2}$ means a half-period translation.
$\hat{\sigma} _{x,y,z}$ are Pauli matrices and $\hat{I_2}$ is the 2$\times$2 identity matrix.
The pseudo-Hermiticity guarantees an associated antilinear symmetry \cite{ALSEigenReal}.
Although the system does not respect $\mathcal{PT}$ symmetry, the bulk antilinear symmetry can instead cancel the gain and loss and give at least partially real spectra.
Meanwhile, the antilinear operation to which $\hat{\mathcal{H}}(k)$ shows the invariance is implicit due to its eigenvector dependence.

$\hat{\mathcal{H}}(k)$ also satisfies a pseudo-anti-Hermiticity \cite{TRSNH,NHedge,Supplementary}, $\hat{\mathcal{H}}(k) = - \hat{\eta} \hat{\mathcal{H}}^{\dagger}(k) \hat{\eta}$, where $\hat{\eta} = \hat{I_2} \otimes \hat{\sigma} _z = {\rm diag}(1,-1,1,-1)$ and $\hat{\eta}^{-1} = \hat{\eta}^{\dagger} = \hat{\eta}$ in our model.
It is known that this symmetry can lead to a nontrivial topology via \textit{chirality} in terms of pairwise eigenvalues, $\omega (k)$ and $-\omega^{*} (k)$.
Since $\hat{\eta}$ is local (diagonal), the resultant topological protection covers all the system parameters.
We notice that the symmetry is equivalent to a particle-hole symmetry, $-\hat{\mathcal{H}}(k) = \hat{\eta} \hat{\mathcal{H}}^{*}(-k) \hat{\eta}$.

\begin{figure}[htbp]
\includegraphics[width=8.6cm]{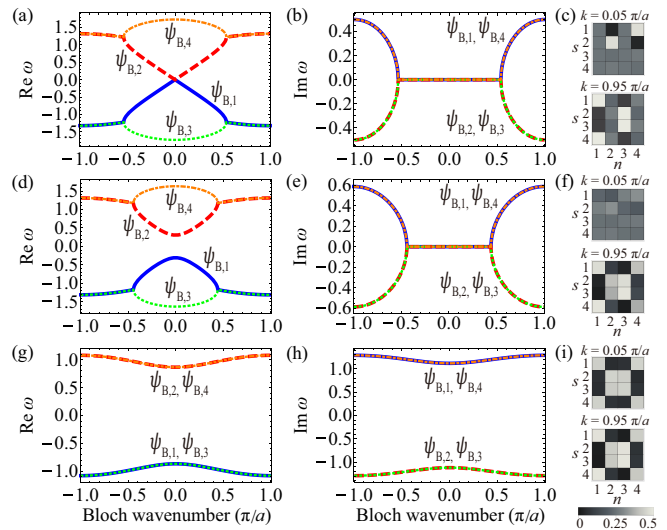}
\caption{(Color online) Band structures and mode patterns of system for different gain and loss profiles. $\kappa = 1$. (a) ${\rm Re} \, \omega (k)$, (b) ${\rm Im} \, \omega (k)$ and (c) $|\psi_{{\rm B},s,n}|^2 / \left\langle \psi _{{\rm B,} s} | \psi _{{\rm B,} s} \right\rangle $ for $g_1 = 1, \ g_2 = 0$, phase (I). (d) ${\rm Re} \, \omega (k)$, (e) ${\rm Im} \, \omega (k)$ and (f) $|\psi_{{\rm B},s,n}|^2 / \left\langle \psi _{{\rm B,} s} | \psi _{{\rm B,} s} \right\rangle $ for $g_1 = 1, \ g_2 = 0.5$, phase (II). (g) ${\rm Re} \, \omega (k)$, (h) ${\rm Im} \, \omega (k)$ and (i) $|\psi_{{\rm B},s,n}|^2 / \left\langle \psi _{{\rm B,} s} | \psi _{{\rm B,} s} \right\rangle $ for $g_1 = 2, \ g_2 = 1$, phase (III). $s$ and $n$: eigenstate and cavity indices.}
\label{fig:band_structure}
\end{figure}

\textit{Bulk properties.--}
Figure \ref{fig:band_structure} shows the real and imaginary band structures and eigenmode distributions for different $g_1$ and $g_2$ values.
When $g _1 < 2$ and $g_2 = 0$ [phase (I) with $A > 0$], Dirac-like dispersion in ${\rm Re} \, \omega (k)$ [Fig. \ref{fig:band_structure} (a)] appears around $\omega (0) = 0$, with cancelled net gain and loss [${\rm Im} \, \omega (k) = 0$, Fig. \ref{fig:band_structure} (b)].
This implies a topological transition point and reflects the antilinear symmetry.
The gapless feature is attributed to $B = 0$ in Eq. (\ref{eq:eigenvalue}).
The band structure also has two EPs meaning the spontaneous antilinear symmetry breaking.
The intensity distributions for $\left| \psi _{{\rm B,} s} \right\rangle $ [Fig. \ref{fig:band_structure} (c)] show that the fields are evenly distributed in the gain and loss cavities ($n = 1, 3$) before this transition ($k = 0.05 \pi/a$), while the eigenmodes with the broken symmetry ($k = 0.95 \pi/a$) exhibit localization at either of them, resulting in complex $\omega (k)$.
Note that $g_1 > 2$ and $g_2 = 0$ [phase (I), $A < 0$] result in a Dirac cone in ${\rm Im} \, \omega (k)$, accompanied by a degeneracy with $\omega (0) = 0$ (Fig. S1 \cite{Supplementary}).

For $g_1, g_2 \ne 0$ under phase (II), a bandgap opens due to $B^2 > 0$ [$|g_2| \le 2 \kappa - g_1$, Fig. \ref{fig:band_structure} (c)], while the real eigenvalues remain around $k = 0$ [Fig. \ref{fig:band_structure} (d)].
Here, the effective couplings between gain and loss cavities become \textit{weaker} than that between the two gain cavities and that between the loss cavities \cite{Supplementary}, as anticipated from the mode localization by $\mathcal{PT}$ symmetry breaking \cite{PTObs09,PTObs10,NHOReview1,NHOReview2}.
This decoupling results in dimerization of the successive cavities with gain and those with loss \cite{Supplementary}.

Mode patterns for the systems with bandgaps confirm this non-Hermitian effect.
Before crossing the EPs in phase (II), the eigenmodes spread over the entire unit cell to cancel the net gain and loss for real $\omega (k)$ [$k = 0.05 \pi/a$, Fig. \ref{fig:band_structure} (f)].
After the symmetry breaking, however, they eventually turn into couples of states with localization at the gain cavities ($n= 1,4$) and loss cavities ($n = 2,3$) for complex $\omega (k)$ ($k = 0.95 \pi/a$).
This feature is clearly distinct from the case of phase (I).
For phase (III), $|g_2| > 2 \kappa - g_1$, the complete antilinear symmetry breaking makes the pairs of upper and lower real bands overlap and gives the split imaginary bands [Fig. \ref{fig:band_structure} (g), (h)].
Thus, only the dimerized eigenstates are allowed all over the Brillouin zone [Fig. \ref{fig:band_structure} (i)].
These data clarify that the non-Hermiticity in our system contributes to the dimerization.
Here, reversing the signs of $g_1$ and $g_2$ does not affect the band structure as seen in Eq. (\ref{eq:eigenvalue}), while a topological transition between systems with $g_2 > 0$ and $g_2 < 0$ is expected at the gap closing with $\omega (0) = 0$.
The non-Hermiticity-based bandgap broadens continuously as $(g_1, g_2)$ gets toward the inside of phase (II) and (III) from their boundaries (Fig. S2 \cite{Supplementary}).

We introduce the normalized global Berry phase \cite{GlobalBP} in $k$-space, $W = \sum_s \frac{i}{4 \pi}\oint dk \langle \left\langle \psi _{{\rm B,} s} \right| \partial _k \left| \psi _{{\rm B,} s} \right\rangle $ ($s = 1, \ldots, 4$), as our topological number.
$W$ denotes the topological feature of the entire system \cite{NABP}, thus it takes account of the problem that the Zak phase \cite{Zakphase} of each band is not discretized in non-Hermitian systems with EPs.
Here, $\left| \psi _{{\rm B,} s} \right\rangle \rangle$ is the left eigenstate that forms a duality with $\left| \psi _{{\rm B,} s} \right\rangle $, namely $\hat{\mathcal{H}}(k) \left| \psi _{{\rm B,} s} \right\rangle = \omega (k) \left| \psi _{{\rm B,} s} \right\rangle$ and $\hat{\mathcal{H}}^{\dagger}(k) \left| \psi _{{\rm B,} s} \right\rangle \rangle = \omega ^* (k) \left| \psi _{{\rm B,} s} \right\rangle \rangle$ \cite{BPBOB}.
The biorthonormal basis $( \{ \left| \psi _{{\rm B,} s} \right\rangle \}, \, \{ \left| \psi _{{\rm B,} s} \right\rangle \rangle \} )$ enables the normalization $\langle \left\langle \psi _{{\rm B,} s} | \psi _{{\rm B,} t} \right\rangle = \delta _{s,t}$ and the extraction of pure geometrical phases from non-Hermitian eigenvectors.
$W$ also reflects the $4 \pi$ periodicity of the eigenvectors \cite{NHAES,Supplementary}.
We obtain integer values of $W = 1$ for $g_2 > 0$ and $W = 0$ for $g_2 < 0$, under $g_1 > 0$ (Fig. S3 \cite{Supplementary}).
$W$ hence confirms the non-Hermiticity-induced nontrivial bulk photonic topology and topological transition between the two conditions in Fig. \ref{fig:schematic} (a).
Interestingly, the discrete change in $W$ also holds in the gapless phases, (IV) and (V).
A geometrical picture of $W$, which illustrates the topological transition, is discussed elsewhere (Fig. S4 and S5 \cite{Supplementary}).

\begin{figure}[htbp]
\includegraphics[width=8.6cm]{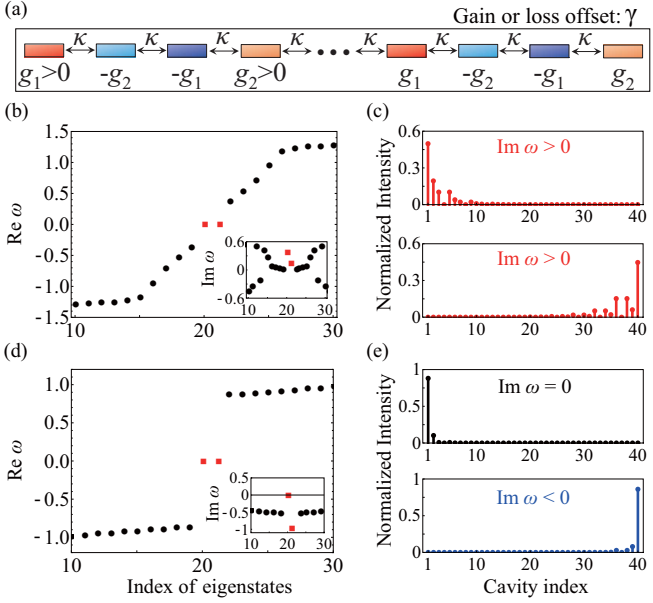}
\caption{(Color online) (a) Illustration of our finite non-Hermitian topological lattice. (b), (d) Sorted and selected real eigenvalues for forty-cavity systems. Insets: corresponding imaginary eigenvalues. Squares: topological edge states. (c), (e) Intensity distributions for the edge states. (b), (c) $g_1 = 1, g_2 = 0.5, \gamma = 0$. (d), (e) $g_1 = 2, g_2 = 1, \gamma \sim -1.569$.}
\label{fig:edge_states}
\end{figure}

\textit{Edge states.--}
Figure \ref{fig:edge_states} shows the topological edge states in our finite systems with 40 cavities for $g_1 > 0, g_2 > 0$.
Here, there is relatively weaker effective coupling between gain and loss cavities at each edge [Fig. \ref{fig:edge_states} (a)], which is similar to the edge-state generation condition in the Hermitian SSH model \cite{SSHModel,PhSSH}.
Displayed eigenfrequencies show a pair of midgap states for both phase (II) [Fig. \ref{fig:edge_states} (b)] and (III) [Fig. \ref{fig:edge_states} (d)].
For Fig. \ref{fig:edge_states} (d), we consider an offset absorption potential $i \gamma$ for every cavity, which only shifts all the eigenfrequencies by $i \gamma$ and cancels ${\rm Im} \, \omega$ of a midgap state.
Each eigenmode with ${\rm Re} \, \omega = 0$ is localized at the left or right edge.
Their ${\rm Im} \, \omega$ reflect the imaginary potential around the relevant edge cavities.
A remarkable difference is whether the localization is unit-based [phase (II), Fig. \ref{fig:edge_states} (c)] or cavity-based [phase (III), Fig. \ref{fig:edge_states} (e)], corresponding to whether the bulk without the imaginary offset is in the \textit{exact} phase [${\rm Im} \, \omega (0) = 0$] or \textit{broken} phase [${\rm Im} \, \omega (0) \ne 0$].
In phase (III), a topological edge mode has the largest ${\rm Im} \, \omega$ and can be the only state that oscillates (${\rm Im} \, \omega = 0$) via the loss offset, $\gamma < 0$ [Fig. \ref{fig:edge_states} (d)].
Note that the edge state are also found when each side is terminated by a loss cavity ($g_1 < 0, g_2 < 0$).
However, they disappear if a cavity on one edge has gain and one on the other undergoes loss ($g_1 g_2 < 0$), as indicated by $W = 0$.
In Supplementary Materials, we show detuned edge states for $g_1 = g_2$, and striking robustness of the edge states to disorder (Fig. S6 and S7 \cite{Supplementary}).
We also find edge states in the gapless phases, and they are explained by $W$ and fractional vortex charges of eigenstates (Fig. S8, S9 and S10 \cite{NHedge,2DAnomoulousTI,Supplementary}).

We can discuss the origin of the topological edge states with ${\rm Re} \, \omega = 0$, in terms of a particle-hole symmetry \cite{TopoClass} equivalent to our pseudo-anti-Hermiticity.
Because our lattice Hamiltonian $\hat{\mathcal{H}}$ with a finite number of cavities and the open boundary condition is a symmetric matrix ($\hat{\mathcal{H}} = \hat{\mathcal{H}}^{\rm T}$), its global pseudo-anti-Hermiticity, $\hat{\mathcal{H}} = - \hat{\eta}' \hat{\mathcal{H}}^{\dagger} \hat{\eta}'$, easily reduces to a particle-hole symmetry, $-\hat{\mathcal{H}} = \hat{\eta}' \hat{\mathcal{H}}^{*} \hat{\eta}'$.
Here, $\hat{\eta}' = {\rm diag}(1,-1,1,-1, \dots, 1,-1,1,-1)$ is again local, $\hat{\eta}'^{-1} = \hat{\eta}'^{\dagger} = \hat{\eta}'$ and $\hat{\mathcal{H}}^{*}$ is the complex conjugate of $\hat{\mathcal{H}}$.
It indicates that the number of states with ${\rm Re} \, \omega = 0$ at each edge can change only by two \cite{PHS1,PHS2}.
Thus, a single isolated edge state with ${\rm Re} \, \omega = 0$ on each side, based on $(1,0,0, \ldots, 0)^{\rm T}$ and $(0, \ldots, 0,0,1)^{\rm T}$ in the system for $g_1 = g_2 = 0$, is topologically protected by this symmetry under proper bandgap-opening conditions \cite{ZeroEn}.

\begin{figure}[htbp]
\includegraphics[width=8.6cm]{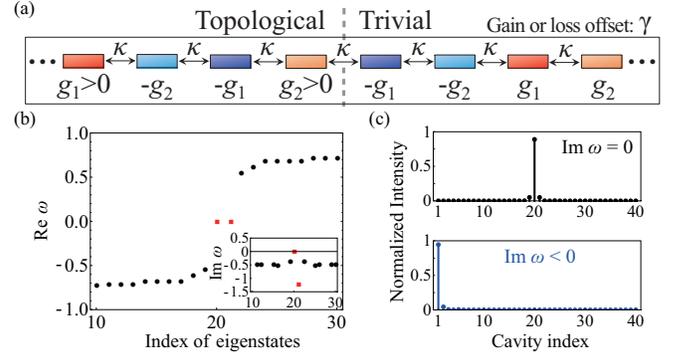}
\caption{(Color online) (a) Interface between topologically nontrivial (left) and trivial (right) lattices, with 20 cavities for each. (b) Sorted and selected ${\rm Re} \, \omega$ of the system. Inset: corresponding ${\rm Im} \, \omega$. Left and right squares: eigenvalues for topological interface and edge states. (c) Intensity profiles for the interface (upper) and edge (lower) modes. $g_1 = 1.5, g_2 = 3, \gamma \sim -2.526$.}
\label{fig:interface_states}
\end{figure}

\textit{Interface states.--}
As an application of our model, we present a controllable non-Hermiticity-based midgap topological interface state (Fig. \ref{fig:interface_states}).
Here, we prepare butting of topologically nontrivial and trivial lattices, effectively forming an ``long-long defect'' at their boundary, as in the SSH model \cite{WATDS} [Fig. \ref{fig:interface_states} (a)].
Both lattices are adjusted to be in phase (III), and larger gain is applied to the interface cavity of the nontrivial array, than its edge ($g_2 > g_1$).
In consequence, a topological interface state, which has ${\rm Re} \, \omega = 0$ and exhibits strong boundary localization, obtains the largest ${\rm Im} \, \omega$ as the abovementioned edge state.
Including a global loss bias $i \gamma$, we can hence expect single-mode lasing (${\rm Im} \, \omega = 0$) of such a state.
The system eigenfrequency profile [Fig. \ref{fig:interface_states} (b)] confirms a pair of midgap states and cancellation of ${\rm Im} \, \omega$ in one of them. 
The intensity distributions of the eigenmodes with $\omega = 0$ and ${\rm Re} \, \omega = 0$, ${\rm Im} \, \omega < 0 $ certainly indicate topological interface and edge states confined at the right ($n = 20$) and left ($n = 1$) sides of the nontrivial lattice, respectively [Fig. \ref{fig:interface_states} (c)].
Both states are topologically protected by the non-Hermitian symmetries.
A topological bound state with global $\mathcal{PT}$ symmetry, which systematically satisfies ${\rm Re} \, \omega = {\rm Im} \, \omega = 0$, can also be demonstrated. (Fig. S11 \cite{Supplementary}).

In conclusion, we have shown that the topological insulating properties of the one-dimensional resonator array can be controlled by the gain and loss.
Our scheme is experimentally feasible by modifying the existing laser arrays with controlled optical pumping \cite{1DTopoLaser1,1DTopoLaser2,1DTopoLaser3} and valid for coupled waveguides \cite{BulkPhTT,BoundPhTS}.
Moreover, it can be explored as an extension of $\mathcal{PT}$-symmetric setups in photonics \cite{PTBHCROW,PTFBLattice,PTUT1,PTUT2,PTLaser1,PTLaser2,PTVCSEL,NHLTopoExp}, phononics \cite{PTPhnlaser,PTAcoustics,PTSBPhn}, and circuit electronics \cite{PTLRCC,UDPTC}.
It would pave the way for various possibilities of non-Hermitian topological photonics, such as reconfigurable topological lasing states, non-Hermiticity-based topological pumping \cite{NHLTopo,TopoPump}, topological superstructures \cite{TopoMultilayer} and Floquet topological systems \cite{DMFTopo,PhFloquetTI}.
Extending the topological controllability to two-dimensional systems is another important direction.

\begin{acknowledgments}
We thank Hiroshi Yamaguchi, Kensuke Inaba and Koji Muraki for fruitful discussions. We acknowledge financial support from the Japan Science and Technology Agency (JST) through the CREST program under Grant No. JPMJCR15N4.
\end{acknowledgments}

\appendix

\section{Gapless Phases}
We have presented a phase diagram for the band structure of the considered system [Fig. 1 (c) in the main text].
Here, we discuss bulk band properties of the systems under gapless conditions.
The gap of ${\rm Re} \, \omega (k)$ closes at the boundary between the phase (III) and (IV), $|g_2| = g_1 - 2 \kappa$ ($k$: Bloch wavenumber, $g_1$, $g_2$: magnitudes of gain and loss, $\kappa$: cavity coupling rate).
Here, linear dispersion in ${\rm Re} \, \omega$ is obtained like the case of phase (I) with $A > 0$ ($g_1 < 2 \kappa$ for $g_2 = 0$).
Because it gives ${\rm Im} \, \omega (0) \ne 0$, however, this boundary is shown to be topologically irrelevant, in terms of $W$ and its geometrical picture.
The band structure in phase (IV) then has a flat region in its real part with ${\rm Re} \, \omega (k) = 0$ around $k = 0$ [Fig. \ref{fig:Band3and4} (a)].
Here, there exist two bifurcation points [exceptional points (EPs)] in ${\rm Re} \, \omega (k)$.
The corresponding curved imaginary bands around $k = 0$ coalesce in pairs at the EPs [Fig. \ref{fig:Band3and4} (b)], forming two \textit{exceptional rings} \cite{ERings} in both ${\rm Im} \, \omega (k) > 0$ and ${\rm Im} \, \omega (k) < 0$.
Such rings are lost in phase (V); the overlapped flat real bands extend [Fig. \ref{fig:Band3and4} (c)], and all the imaginary band curves are split [Fig. \ref{fig:Band3and4} (d)] over the entire first Brillouin zone.
We notice that two eigenstates in phase (IV) [Fig. \ref{fig:Band3and4} (b)] have jumps in ${\rm Im} \, \omega (k)$ at EPs.
This is because $\omega (k)$ has double radical signs, and such jumps are caused by non-Hermiticity.

When in phase (I) with $A < 0$ ($g_1 > 2 \kappa$ for $g_2 = 0$), the band structure shows a Dirac cone in ${\rm Im} \, \omega (k)$, instead of ${\rm Re} \, \omega (k)$ [Fig. \ref{fig:Band3and4} (e) and (f)].
As in phase (I) with $A > 0$ [Fig. 2 (a) and (b)], the system here has a degeneracy with $\omega (0) = 0$ and a pair of distinct eigenmodes.
Such a degenerate point for $\omega = 0$ is later clarified to be the indication of the topological transition point in our non-Hermitian system, as well as Hermitian topological systems.
It is noteworthy that the Dirac cone in ${\rm Im} \, \omega (k)$ with the zero-detuning degeneracy also remains for the case of larger $g_1$, including the points $(g_1, g_2)$ in contact with phase (V).

\section{Bulk symmetry for nontrivial topology}
The bulk system should have a symmetry that results in its nontrivial topology with the finite $W$.
\begin{figure}[htbp]
\includegraphics[width=8.6cm]{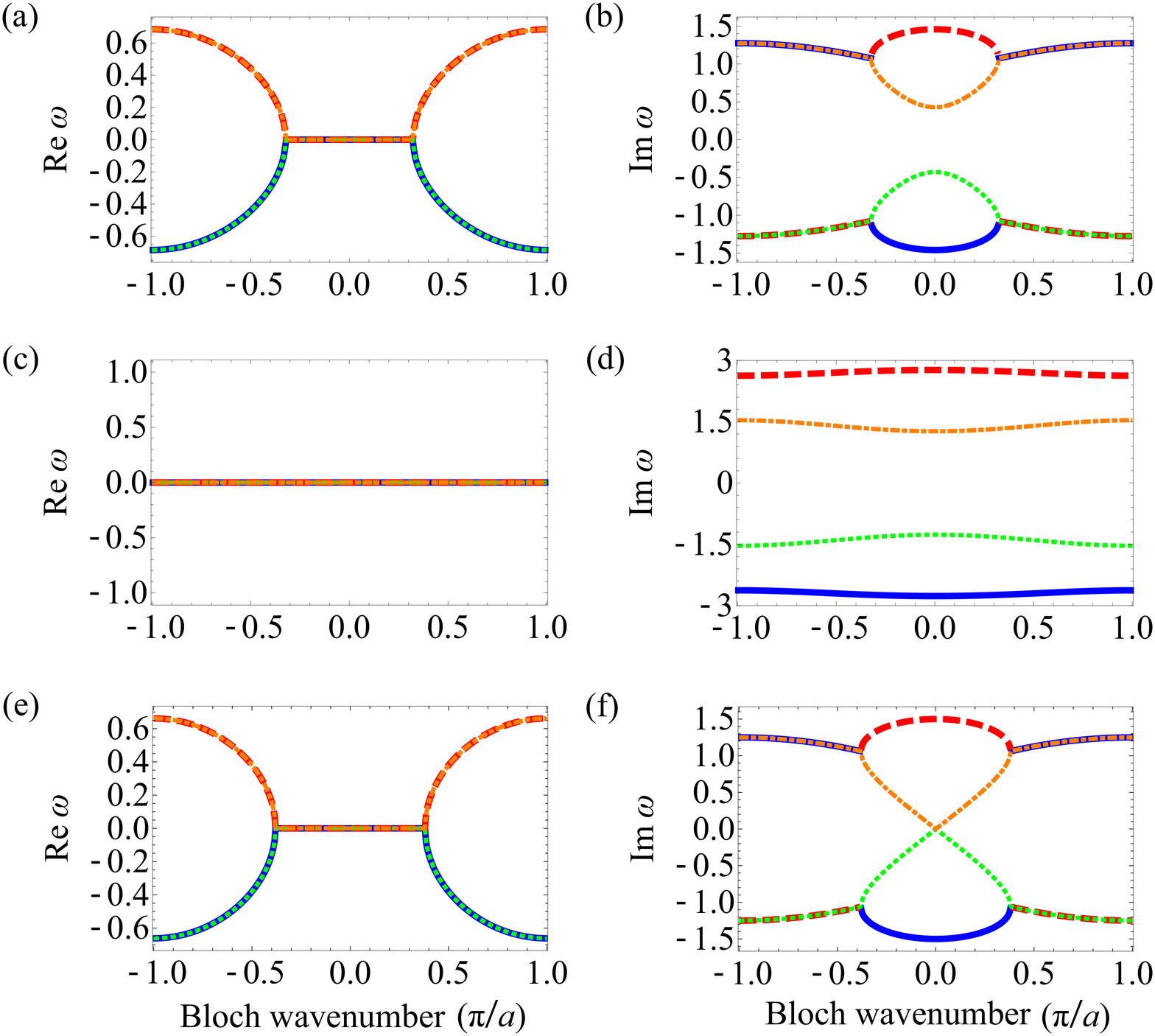}
\caption{Real and imaginary band structures of the system in the gapless phases. $\kappa = 1$. (a) ${\rm Re} \, \omega$ and (b) ${\rm Im} \, \omega$ for $g_1 = 2.5$ and $g_2 = 0.25$, in phase (IV). (c) ${\rm Re} \, \omega$ and (d) ${\rm Im} \, \omega$ for $g_1 = 3.5$ and $g_2 = 1$, in phase (V). (e) ${\rm Re} \, \omega$ and (f) ${\rm Im} \, \omega$ for $g_1 = 2.5$ and $g_2 = 0$, in phase (I) with $A < 0$.}
\label{fig:Band3and4}
\end{figure}
Here, $\hat{\mathcal{H}}(k)$ has a pseudo-anti-Hermiticity \cite{NHedge}, $\hat{\mathcal{H}}(k) = - \hat{\eta} \hat{\mathcal{H}}^{\dagger}(k) \hat{\eta}$, where $\hat{\eta} = {\rm diag}(1,-1,1,-1)$ in our system.
The symmetry supports pairs of states with eigenfrequencies of $\omega$ and $-\omega^{*}$ via the duality relation, for the case of ${\rm Re} \, \omega \ne 0$.
When the eigenstates for each $k$ in the gapless phases have ${\rm Re} \, \omega = 0$ and different ${\rm Im} \, \omega$, as seen in Fig. \ref{fig:Band3and4}, their eigenvalues cannot satisfy the pairing relation ($\omega$, $-\omega^{*}$).
We later discuss a chirality of the states based on this absence of pairwise right eigenstates for ${\rm Re} \, \omega = 0$.

$\hat{\mathcal{H}}(k)$ also has a chiral symmetry $\hat{\mathcal{C}}(k) \hat{\mathcal{H}}(k) \hat{\mathcal{C}}(k) = -\hat{\mathcal{H}}(k)$ with $\hat{\mathcal{C}}(k) = \hat{\mathcal{S}}(k) \hat{\eta}$ and $\hat{\mathcal{S}}(k) = \hat{\sigma} _x \otimes (\cos \frac{k}{2}) \hat{I_2} + \hat{\sigma} _y \otimes (\sin \frac{k}{2}) \hat{I_2}$, which guarantees couples of eigenvalues $\pm \omega (k)$.
When $g_1 \ne 0, g_2 = 0$ or $g_1 =0, g_2 \ne 0$, the bulk system always has two degenerate states with ${\rm Re} \, \omega = {\rm Im} \, \omega = 0$ at $k = 0$, regardless of the value of the finite gain and loss magnitude, $g_1$ or $g_2$.
Here, $\omega (k)$ forms a Dirac cone around ${\rm Re} \, \omega = 0$ [Fig. 2 (a), $|g_1| < 2, g_2 = 0$ or $g_1 =0, |g_2| < 2$] or ${\rm Im} \, \omega = 0$ [Fig. \ref{fig:Band3and4} (f), $|g_1| > 2, g_2 = 0$ or $g_1 =0, |g_2| > 2$], depending on the parameters.
One exceptional case (not plotted here) is $g_1 = 2, g_2 = 0$ or $g_1 =0, g_2 = 2$, where coalesced exceptional points are formed at both ${\rm Re} \, \omega = 0$ and ${\rm Im} \, \omega  = 0$ at $k = 0$.
These degeneracies with $\omega = 0$ can be considered as an indication of the bulk chiral symmetry, or highly symmetric points in terms of pseudo-anti-Hermiticity (with the help of antilinear symmetry).

The system band structures, composed of four eigenstates, then satisfy both pseudo-anti-Hermiticity and chiral symmetry.
Here, the pseudo-anti-Hermiticity can contribute to the topological protection in terms of $\kappa$, $g_1$ and $g_2$, because it is based on the purely local operator, $\hat{\eta} = \hat{\eta}^{-1} = \hat{\eta}^{\dagger}$.
Moreover, the bulk pseudo-anti-Hermiticity is equivalent to the particle-hole symmetry with the same local operator, $-\hat{\mathcal{H}}(k) = \hat{\eta} \hat{\mathcal{H}}^{*}(-k) \hat{\eta}$.
Meanwhile, since $\hat{\mathcal{C}}(k)$ includes the translation operator $\hat{\mathcal{S}}(k)$, the chiral symmetry is missing in our finite chains with termination at both sides.
Thus, the bulk chiral symmetry cannot by itself explain the topological robustness of the edge states.

\section{Effective Decoupling by Gain and Loss}
We can estimate the effective cavity decoupling induced by gain and loss with a simple two-cavity system. We consider the Hamiltonian,
\begin{equation}
\hat{\mathcal{H}}_2 = \left( \begin{array}{cc}
i \gamma_1 & \kappa  \\
 \kappa & i \gamma_2  \\
\end{array} \right), \label{eq:twocavHamiltonian}
\end{equation}
and its eigenvalues $\lambda = i(\gamma_1 + \gamma_2)/2 \pm \sqrt{\kappa^2 - (\gamma_1 - \gamma_2)^2/4}$, where $\gamma_1$ and $\gamma_2$ are the imaginary potentials of the cavities.
By comparing $\lambda$ with the eigenvalues without any gain and loss, $\lambda(g_1, g_2 = 0) = \pm \kappa$, we can understand that the coupling, in terms of the splitting of ${\rm Re} \, \lambda$, is effectively reduced from $\kappa$ to $\kappa' = \sqrt{\kappa^2 - (\gamma_1 - \gamma_2)^2/4}$.
Thus, it is expected that the local effective coupling decreases depending on the difference between the imaginary potentials of adjacent cavities, $|\gamma_1 - \gamma_2|$.

With this supposition, the four-cavity system considered in this work provides two effective couplings, $\kappa_1' = \sqrt{\kappa^2 - (g_1 + g_2)^2/4}$ and $\kappa_2' = \sqrt{\kappa^2 - (g_1 - g_2)^2/4}$.
Here, we can see that substituting these effective couplings into the bandgap of the Hermitian SSH model \cite{SSHModel} reproduces that of our four-cavity model, $\Delta = \sqrt{2} \sqrt{A - \sqrt{A^2 - B^2}}$, where $A = 4 \kappa^2 - g_1 ^2 - g_2 ^2$ and $B = 2 g_1 g_2$.
This implies that the gradient of the system's imaginary potential is relevant to dimerization, as discussed in the main text.
This correspondence is because, at the band center giving the frequency gap ($k = 0$), the impact of periodicity is eliminated and the pseudo-Hermiticity might cancel out the effect of ${\rm Im} \, \lambda$.
It is noteworthy that the SSH model does not reproduce the band structure of the four-cavity system for $k \ne 0$.

Here, we emphasize that the non-Hermitian dimerization based on such effective decoupling is a consequence of using the unit cell of four cavities.
Our system is considered as a minimum non-Hermitian extension of Hermitian two-cavity units, based on doubling the number of cavities for introducing the pairs of gain and loss $(\pm g_1, \pm g_2)$ to each cavity.

\section{Bandgap}
\begin{figure}[htbp]
\includegraphics[width=6.0cm]{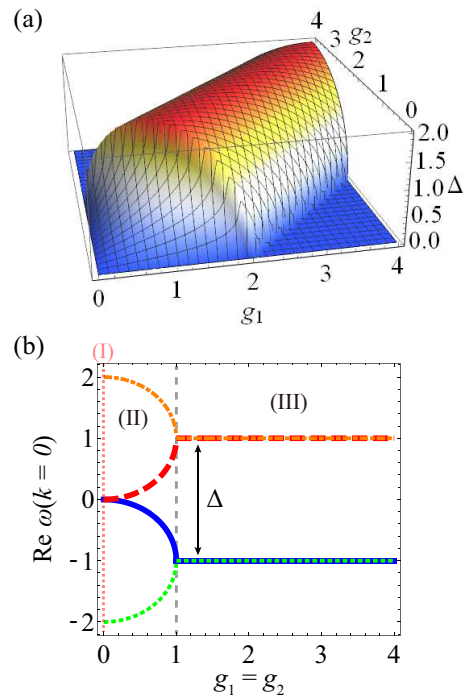}
\caption{The system frequency bandgap $\Delta$ dependent on the gain and loss. (a) $\Delta$ for the two-dimensional parameter space $(g_1 > 0, g_2 > 0)$. (b) ${\rm Re} \, \omega (k=0)$ for the four eigenstates depending on $g_1 = g_2 \ (g_1 \ge 0)$, for phase (I) ($g_1 = 0$), (II) ($0 < g_1 < 1$) and (III) ($g_1 > 1$). $\Delta$ is created by the non-Hermitian band modulation and EP formation.}
\label{fig:Bandgap}
\end{figure}

Figure \ref{fig:Bandgap} shows the width of the system bandgap $\Delta$ for ${\rm Re} \, \omega (k)$ depending on the gain and loss.
Here, the bandgap in our model is determined by the eigenfrequencies at $k = 0$.
Fig. \ref{fig:Bandgap} (a) presents the dependence of $\Delta$ in the two-dimensional parameter space $(g_1 > 0, g_2 > 0)$.
$\Delta$ enlarges continuously as $g_1$ and $g_2$ increase in phase (II).
In contrast, the gap in phase (III) has the maximum ($\Delta = 2$) along with $g_1 = g_2$ and falls down to zero abruptly at the boundary between phase (III) and (IV), by the rise in either $g_1$ or $g_2$.

To see how the bandgap opens, we depict the real eigendetuning profile at the band center ${\rm Re} \, \omega (k = 0)$, depending on the gain and loss [Fig. \ref{fig:Bandgap} (b)].
Here, both $g_1$ and $g_2$ are varied so that their values are the same $g_1= g_2 \ (g_1 \ge 0)$, and $\Delta$ corresponds to the difference between the second and third largest real eigendetuning.
When $g_1$ increases from zero, the parameters $(g_1, g_2)$ vary in parallel with the boundary between phase (III) and (IV), away from $(0,0)$ on phase (I).
As a result, the degeneracy at ${\rm Re} \, \omega (0) = 0$ gently splits into two branches, which form the bandgap.
Each of the upper and lower pairs of branches then sharply coalesces at $g_1 = g_2 = 1$, resulting in $\Delta = \sqrt{2(4\kappa ^2 - g_1 ^2 -g_2 ^2)} = 2$.
This shows that the bandgap is based on the non-Hermitian band modulation and EP formation.
The coalescence (EP) at $g_1 = g_2 = 1$ reflects the complete antilinear symmetry breaking at the boundary between phase (II) and (III).
$\Delta$ is then constant in phase (III) for this dependence ($g_1 = g_2$, $g_1 > 1$), meaning that the dimers keep a constant effective coupling that depends on $|g_1 - g_2|$.

\section{Topological number}
Our topological invariance $W$, the normalized global Berry phase \cite{GlobalBP}, is the trace of the single-parameter non-Abelian Berry phase matrix in terms of the biorthonormal basis.
This quantifies the topology of the whole non-Hermitian band structure and is applicable to systems with degeneracies.
\begin{figure}[htbp]
\includegraphics[width=6.5cm]{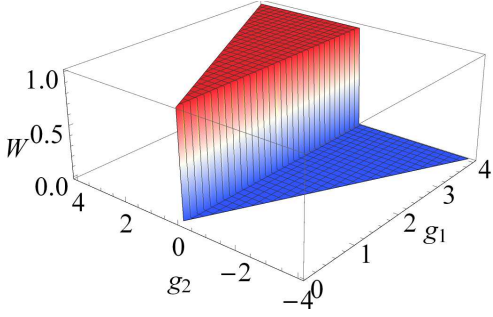}
\caption{Global Berry phase $W$ for different $g_1$ and $g_2$ values, which cover the gapped and gapless phases. $W$ shows a discontinuous change: $W = 1$ for $g_2 > 0$ and $W = 0$ for $g_2 < 0$, showing the topological transition at the gap closing, $g_2 = 0$. It can be said that ${\rm Im} \, W$ is numerically zero. The assigned differential step of wavenumber is $\Delta k = 10^{-10} \pi /a$. Steps of calculation points in ($g_1$, $g_2$) are both 0.05. For the case of the transition point ($g_2 = 0$), $W = 1/2$, and this is also seen in the Hermitian and non-Hermitian SSH systems.}
\label{fig:BerryPhase}
\end{figure}
Figure \ref{fig:BerryPhase} shows $W$ of the system in both the gapped and gapless conditions under $g_1 >0$.
Here, $W = 1$ for $g_2 > 0$ and $W = 0$ for $g_2 < 0$ are confirmed with negligible errors, for all the considered values of $(g_1, g_2)$.
It confirms the non-Hermitian topological transition at $g_2 = 0$, where the Dirac cone in ${\rm Re} \, \omega (k)$ or ${\rm Im} \, \omega (k)$ emerges (except for $g_1 = 2$).
Note that the global Berry phase avoids the difficulty of the 4$\times$4 Hamiltonian not giving off-block-diagonal $Q$ matrices in a standard construction of the winding number \cite{NHedge}.

It is noteworthy that the two EPs encountered in phase (II) and (IV) are excluded from the integral range in the $W$ calculation, because of the following special properties of EPs.
First, each EP switches the pairing of the dual analytic $\left| \psi _{{\rm B,} s} \right\rangle$ and $\left| \psi _{{\rm B,} s} \right\rangle \rangle$ due to the emergence of finite ${\rm Im} \, \omega (k)$ \cite{PhysNHD}.
Thus, numerical integration cannot involve an EP when a single left eigenstate $\langle \left \langle \psi_{{\rm B,} s} \right|$ is considered.
Second, it is known that the norm $\langle \langle \psi_{{\rm B,} s} | \psi_{{\rm B,} s} \rangle$ itself vanishes at EPs (self-orthogonalization) \cite{PhysEP}.
This makes unable for numerical differentiation algorithms to use the eigenstates exactly at EPs, even under the normalization based on the biorthonormal basis.
Note that we took into consideration the combined EP in ${\rm Im} \, \omega$ remaining at $k = 0$ for the boundary between phases (II) and (III).
The removed wavenumber ranges are $[-k_{\rm EP} - 2 \Delta k, -k_{\rm EP} + 2 \Delta k]$ and $[k_{\rm EP} - 2 \Delta k, k_{\rm EP} + 2 \Delta k]$, where $k_{\rm EP}$ denotes an EP with $k>0$.
Fortunately, these widths can be arbitrarily small, depending on the unit wavenumber step $\Delta k$ in the numerical differential.
We also find that the eigenvectors in phase (IV) have peculiar divergence of their elements at $k = \pm \pi /a$, so these points have to be removed in the same way. 
We finally obtain $W \rightarrow 1$ for $g_2 > 0$, $W \rightarrow 0$ for $g_2 < 0$ and $W \rightarrow 1/2$ for $g_2 = 0$ with $\Delta k \rightarrow 0$ [Fig. \ref{fig:BerryPhase} (a)].
Here, $\Delta k = 10^{-10} \pi /a$ is used for the computation.
$|{\rm Im} \, W|$ values for all $(g_1, g_2)$ points are less than $10^{-7}$ and hence are numerically zero.
Note that there is no constraint above in phase (III) and (V).
The biorthonormality of the eigenstates is guaranteed except for the EPs, because the states are nondegenerate \cite{NHSE}.
The eigenstates in phase (III) and (V) with coalesced real bands have split imaginary bands, thus we surely see that they are not degenerate or problematic in computing $W$.

We further notice that the eigenstates are $4\pi$-periodic in $k$ by non-Hermiticity, as pointed out in Ref. \onlinecite{NHAES}.
The loop of $k$ for the states should therefore be two rounds of the first Brillouin zone, and the winding number should be an average of the Berry phase per round \cite{ReplyLee}, namely,
\begin{equation}
W = \sum_s \frac{i}{2 \pi} \left( \frac{1}{2} \int_{-2\pi}^{2\pi} dk \langle \left\langle \psi _{{\rm B,} s} \right| \partial _k \left| \psi _{{\rm B,} s} \right\rangle \right). \label{eq:TRW}
\end{equation}
Although, we point out that the Berry connection in this case is $2\pi$-periodic and that the Berry phase accumulated in a single round of the Brillouin zone equals Eq. (\ref{eq:TRW}).
This indicates that two EPs are encircled in the parameter space of the effective Hamiltonian.
We show such a geometrical interpretation of $W$ in the next section.

\section{Geometrical picture of topological number}
The topological number $W$ is based on the complex Zak phase in one dimension.
Thus, it could be reduced to the geometrical winding \cite{ZakPhaseGraphene,NHAES,WNNHSSH} of $k$-dependent Hamiltonian parameters around the degeneracies at the origin of the symmetry-protected edge states, i.e. $\omega = 0$.
With the Pauli matrices, our Hamiltonian is written as,
\begin{widetext}
\begin{eqnarray}
\hat{\mathcal{H}}(h_x, h_y) &=& \left( \begin{array}{cccc}
i g_1 & \kappa & 0 & \kappa (h_x - i h_y)^2 \\
 \kappa & -i g_2 & \kappa ({h_x}^2 + {h_y}^2) & 0 \\
0 & \kappa ({h_x}^2 + {h_y}^2) & -i g_1 & \kappa \\
\kappa (h_x + i h_y)^2 & 0 & \kappa & i g_2 \\
\end{array} \right) \nonumber \\
&=& \kappa \left( h_x  \hat{\sigma} _x + h_y  \hat{\sigma} _y \right) \otimes \left( h_x  \hat{\sigma} _x + h_y  \hat{\sigma} _y \right) + \kappa \hat{I_2} \otimes \hat{\sigma} _x + \frac{g_1 + g_2}{2} \hat{\sigma} _z \otimes \hat{\sigma} _z + \frac{g_1 - g_2}{2} \hat{I_2} \otimes \hat{\sigma} _z, \label{eq:hxhyHamiltonian}
\end{eqnarray}
\end{widetext}
where $h_x = \cos (k a/2)$, $h_y = \sin (k a/2)$ and ${h_x}^2 + {h_y}^2 = 1$ for Eq. (1) in the main text.
We see that the two copies of pseudospins form unit circles in the $(h_x, h_y)$ plane, in terms of $\hat{\sigma} _x$ and $\hat{\sigma} _y$.
Considering the doubled period for introducing gain and loss and non-Hermitian effects, it will be appropriate to interpret the winding number as the half of the number of the degenerate points (DPs) for $\omega = 0$ that are enclosed by the counter-clockwise loop of $(h_x, h_y)$, from $k = 0$ to $4 \pi$ \cite{NHAES} ($\mathbb{Z}/2$).
Compared to other possible parameterization based on $\{\cos k a, \sin k a\}$, the formulation above is unique in that it involves all the antidiagonal elements of the Hamiltonian and hence zero-detuning degeneracies based on dimerization conditions.

\begin{figure}[htbp]
\includegraphics[width=8.6cm]{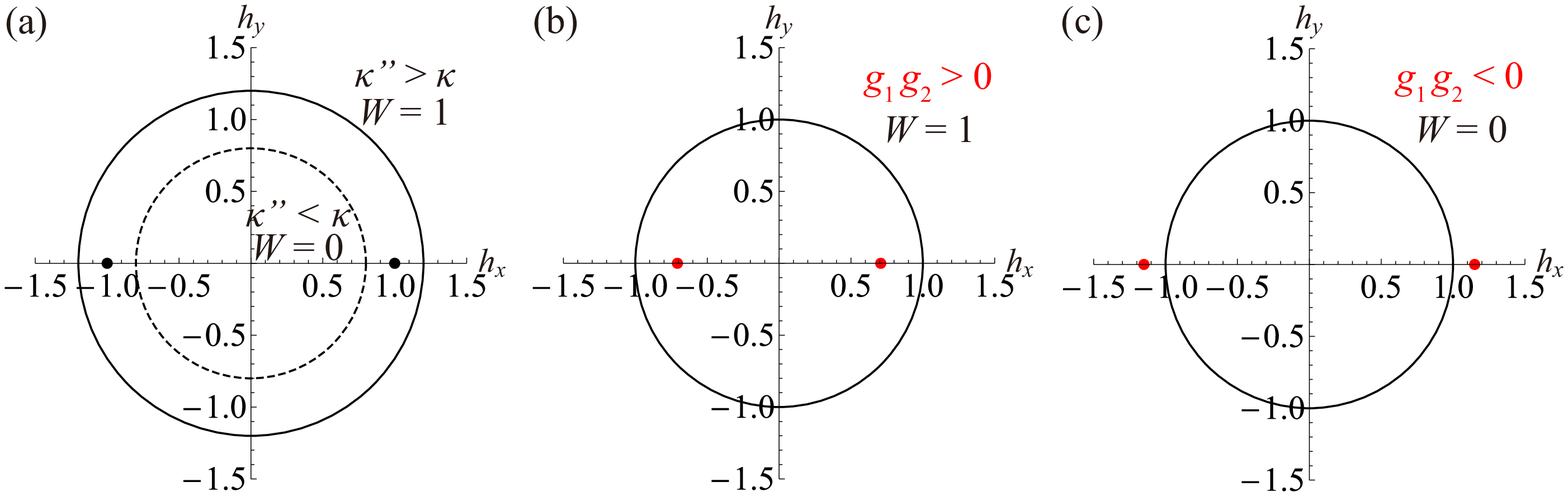}
\caption{Geometrical picture of the winding number $W$ in the system around the topological transition. (a) SSH model, where the two DPs for $\omega = 0$ at $(h_x, h_y) = (\pm 1, 0)$ are in and out of the loop of the coupling parameters, $\kappa'' > \kappa$ and $\kappa'' < \kappa$ meaning $W = 1$ and $W = 0$, respectively. (b), (c) The considered system with gain, loss and the uniform cavity coupling, in phase (II). Here, the DPs are moved to $(h_x, h_y) = (\pm (1 - g_1 g_2/\kappa ^2)^{1/4}, 0)$ by the gain and loss (red dots). $\kappa =1$. (b) The DPs with $g_1 = 1$, $g_2 = 0.75$ are encircled by the fixed loop ${h_x}^2 + {h_y}^2 = 1$, resulting in $W = 1$. (c) Those with $g_1 = 1$, $g_2 = -0.75$ are out of the winding and hence $W = 0$.}
\label{fig:EPs}
\end{figure}

Eq. (\ref{eq:hxhyHamiltonian}) with $g_1 = g_2 = 0$ can cover the SSH model with the four-cavity period by allowing the scaling of the loop, $h_x = h_a \cos (k a/2)$, $h_y = h_a \sin (k a/2)$ and $h_a > 0$.
Here, because of the term $\kappa \hat{I_2} \otimes \hat{\sigma} _x$, the DPs in the $(h_x, h_y) \in \mathbb{R}^2$ space for the Hermitian system are located at $(h_x, h_y) = (\pm 1, 0)$ [Fig. \ref{fig:EPs} (a)].
Except for the Bloch phase factors $\exp (\pm i k a)$, the net coupling between the first and fourth cavities and that between the second and third are $\kappa'' \equiv \kappa ({h_x}^2 + {h_y}^2) = \kappa {h_a}^2 > 0$ ($\kappa > 0$ assumed here).
Thus, when ${h_a}^2 < 1$, i.e. $\kappa'' < \kappa$, the DPs are out of the winding of the coupling parameters.
Meanwhile, a larger loop with ${h_a}^2 > 1$ ($\kappa'' > \kappa$) comes to enclose both DPs.
These cases match rightly the topologically trivial and nontrivial phases of the SSH model, respectively.
This indicates that the nontrivial Berry charge corresponds to the winding of the DPs for $\omega = 0$ in the $(h_x, h_y)$ plane.

While the SSH model scales the loop of the coupling parameters, our model moves the DPs via gain and loss parameters, $g_1$ and $g_2$.
Since the Dirac point with $\omega = 0$ of the system band structures for $g_2 = 0$ or $g_1 = 0$ appears at $k = 0$, we can omit the imaginary ($\hat{\sigma} _y$) term,  $h_y = 0$.
The resultant condition for the degeneracy is given by,
\begin{align}
- &g_1^2- g_2^2 + 2 (\kappa ^2 + {\kappa''} ^2) \nonumber \\ 
&+ \sqrt{ [(g_1 + g_2)^2 - 4 \kappa ^2] [(g_1 - g_2)^2 - 4 {\kappa''} ^2]  }  = 0,
\end{align}
and the solution for $\kappa'' > 0$  is,
\begin{equation}
\kappa'' = \kappa {h_x}^2 = \sqrt{\kappa ^2 - g_1 g_2},
\end{equation}
for $\kappa ^2 \ge g_1 g_2$. Therefore, by applying the gain and loss, the DPs are displaced to,
\begin{equation}
(h_x, h_y) = \left(\pm \left(1 - \frac{g_1 g_2}{\kappa ^2} \right)^\frac{1}{4}, \, 0 \right),
\end{equation}
as shown in Fig. \ref{fig:EPs} (b) and (c).
Our system considers the uniform cavity couplings, namely ${h_x}^2 + {h_y}^2 = 1$.
Thus, the two DPs are encircled by the loop for the case of $g_1 g_2 > 0$ [Fig. \ref{fig:EPs} (b)]; otherwise, they are placed outside [$g_1 g_2 < 0$, Fig. \ref{fig:EPs} (c)].
Again, these conditions are consistent with the topological transition at $g_1 g_2 = 0$ that includes the gapless phases, between $W = 1, g_2 > 0$ and $W = 0, g_2 < 0$ under $g_1 > 0$ in our work.
It is noteworthy that $W$ can only change by unity via the transition, because the system does not have any faster winding factors like second-nearsest-neighbor couplings and the two DPs are located symmetrically to $(h_x, h_y) = (0, 0)$ \cite{WNNHSSH}.
\begin{figure}[htbp]
\includegraphics[width=8.6cm]{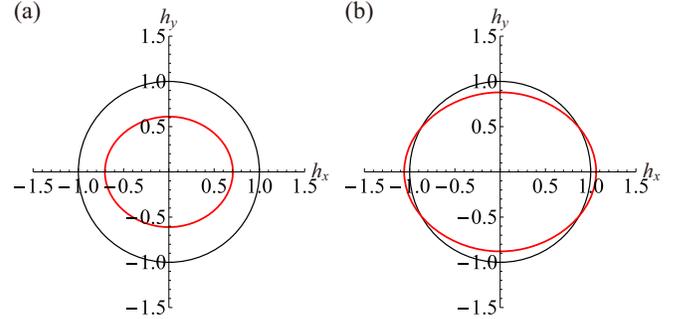}
\caption{Breakdown of geometrical picture of $W$ as the winding of two DPs, for the case of $g_1 g_2 > \kappa ^2$. After the two DPs coalesce and vanish at the origin, degeneracies with ${\rm Re}\, \omega = 0$ are numerically found and form a closed loop (red curve) in the $(h_x, h_y) \in \mathbb{R}^2$ space. They are not always limited to the inside of the loop of the coupling parameters, ${h_x}^2 + {h_y}^2 = 1$ (black curve). (a) $g_1 = 2$, $g_2 = 1$, in phase (III). (b) $g_1 = 2.75$, $g_2 = 0.5$, in phase (IV).}
\label{fig:LODs}
\end{figure}

When $g_1 g_2 < 0$ ($W = 0$), the DPs for $\omega = 0$ always exist and are out of the loop in the $(h_x, h_y) \in \mathbb{R}^2$ plane.
In contrast, the conditions $\kappa ^2 \ge g_1 g_2$ and $g_1 g_2 > 0$ ($W = 1$) only cover phase (I), (II), and small portions of phase (III), (IV) and (V).
As $|g_1|$ and $|g_2|$ increase, the pair of DPs vanish at $(h_x, h_y) = (0, 0)$ with $g_1 g_2 = \kappa ^2$.
When $g_1 g_2 > \kappa ^2$, we can only obtain non-Hermitian degeneracies with ${\rm Re}\, \omega = 0$ and ${\rm Im}\, \omega \ne 0$, satisfying the condition,
\begin{align}
- (g_1^2 &- g_2^2)^2 \,[(g_1 + g_2)^2 - 4 \kappa ^2] \nonumber \\ 
 - & 4 \kappa ^2 \, [(g_1 + g_2)^2 - 2 \kappa ^2] \, (h_x ^2 + h_y ^2)^2 \nonumber \\ 
   &+ 8 \kappa ^4 \, (h_x ^2 + h_y ^2)^2 \, (h_x ^2 - h_y ^2)^2 = 0.
\end{align}
As shown in Fig. \ref{fig:LODs}, however, such degeneracies are no longer independent points but trace a closed loop in the parameter space.
We numerically find that they stay inside of the trajectory of the coupling (${h_x}^2 + {h_y}^2 = 1$) for phase (III) [Fig. \ref{fig:LODs} (a)], while they can go out of it for the gapless phases [Fig. \ref{fig:LODs} (b)]. 
This result indicates the complete antilinear symmetry breaking induced by gain and loss.
Overall, the standard geometrical picture of the winding number covering the SSH model breaks down in the broad parameter region, $g_1 g_2 > \kappa ^2$ in our model.
Although $W$ has been successfully determined in our work, unveiling the source of the Berry charge will be an important direction.

\section{Detuned edge states}
\begin{figure}[htbp]
\includegraphics[width=8.6cm]{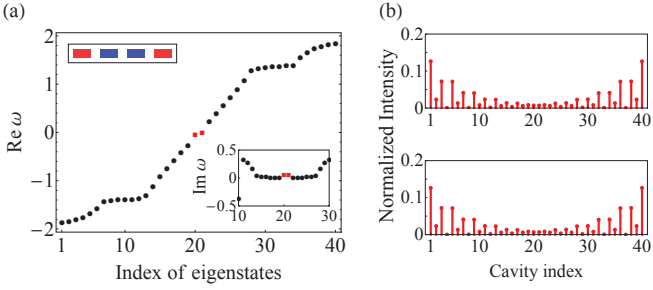}
\caption{Eigenstates for $g_1 = g_2 = 0.5$, of a 40 cavity system. (a) Sorted ${\rm Re} \, \omega$. Upper left inset: schematic of the unit cell, with an imaginary potential of $(0.5 i, -0.5 i, -0.5 i, 0.5 i)$. Lower right inset: ${\rm Im} \, \omega$ for selected states. Squares: midgap states with ${\rm Re} \, \omega \ne 0$. (b) Intensity distributions for the midgap states, with remaining weak couplings between the left-half and right-half edge modes.}
\label{fig:MidgapStates}
\end{figure}

When $g_1 = g_2$, the behavior of the midgap edge states is different from those shown in the main text.
For the system in phase (II), the real parts of their eigenfrequencies are slightly detuned from the single cavity resonance, i.e. ${\rm Re} \, \omega \ne 0$ [Fig. \ref{fig:MidgapStates} (a)].
Their net gain ${\rm Im} \, \omega \ne 0$ is relatively small compared to systems with similar $(g_1, g_2)$ satisfying $g_1 \ne g_2$.
The mode patterns of these states show the intensity accumulation at both sides and their tails are extended to the center of the lattice [Fig. \ref{fig:MidgapStates} (b)].
This real detuning decreases as $g_1$ and $g_2$ increase, and vanishes in phase (III).

Such peculiar properties stem from the interplay between the bulk antilinear symmetry and global mirror symmetry of the finite system.
Because the eigenstates of the inversion operator $\hat{P}$ must have even or odd parity with respect to the center of the lattice, the mirror symmetry $[\hat{\mathcal{H}},\hat{P}]=0$ localizes the midgap modes at both edges.
Moreover, the exact antilinear symmetry of the Bloch eigenmodes around $k = 0$ in phase (II) does not provide any sublattice mode localization, thus somewhat extended states are allowed in finite systems.
Consequently, while the topological charge sweeps photons towards the edges, the right-half and left-half cavity cluster modes are not completely decoupled.
The right and left cluster modes are in-phase and out-of-phase for the midgap states with lower and higher ${\rm Re} \, \omega$, respectively.
Meanwhile, we find that increasing the system cavity number, with retaining $g_1$ and $g_2$ values, barely affects the decay profile of the edge modes and hence suppresses the detuning (coupling between the edges).
It means that these features are finite-size effects, and edge states with ${\rm Re} \, \omega = 0$ are restored in the limit of the infinite system size \cite{ZeroEn}.

\section{Robustness of topological edge states}
The edge states of the finite chains are topologically protected by the pseudo-anti-Hermiticity and equivalent particle-hole symmetry.
Here, because these symmetries root on a local operator, no fluctuation in any of the parameters considered in the model can break them.
Therefore, the edge states retain ${\rm Re} \, \omega = 0$ under such perturbation.
To demonstrate this, we modify the lattice Hamiltonian as follows,
\begin{widetext}
\begin{equation}
\hat{\mathcal{H}_{\rm D}} = \left( \begin{array}{cccccc}
\ddots & \kappa \xi_{\kappa,4 l - 4} & 0 & 0 & 0 & 0  \\
\kappa \xi_{\kappa,4 l - 4} & i g_1 \xi_{g,4 l - 3} & \kappa \xi_{\kappa,4 l - 3} & 0 & 0 & 0 \\
0 & \kappa \xi_{\kappa,4 l - 3} & -i g_2 \xi_{g,4 l - 2} & \kappa \xi_{\kappa,4 l - 2} & 0 & 0 \\
0 & 0 & \kappa \xi_{\kappa,4 l - 2} & -i g_1 \xi_{g,4 l  - 1} & \kappa \xi_{\kappa,4 l - 1} & 0 \\
0 & 0 & 0 & \kappa \xi_{\kappa,4 l - 1} & i g_2 \xi_{g,4 l} & \kappa \xi_{\kappa,4 l} \\
0 & 0 & 0 & 0 & \kappa \xi_{\kappa,4 l} & \ddots \\
\end{array} \right), \label{eq:Hamiltonian2}
\end{equation}
\end{widetext}
where $l$ is the unit index and all the parameters are real numbers.
The fluctuation coefficients $\{\xi_{\kappa, n}\}$ and $\{\xi_{g, n}\}$ ($n$: cavity index) are all independent random numbers under Gaussian distribution with a mean of unity and a standard deviation of $\sigma$.
Note that $\hat{\mathcal{H}_{\rm D}}$ is assumed to be a symmetric matrix due to the reciprocity of Hermitian cavity couplings.

\begin{figure}[htbp]
\includegraphics[width=8.6cm]{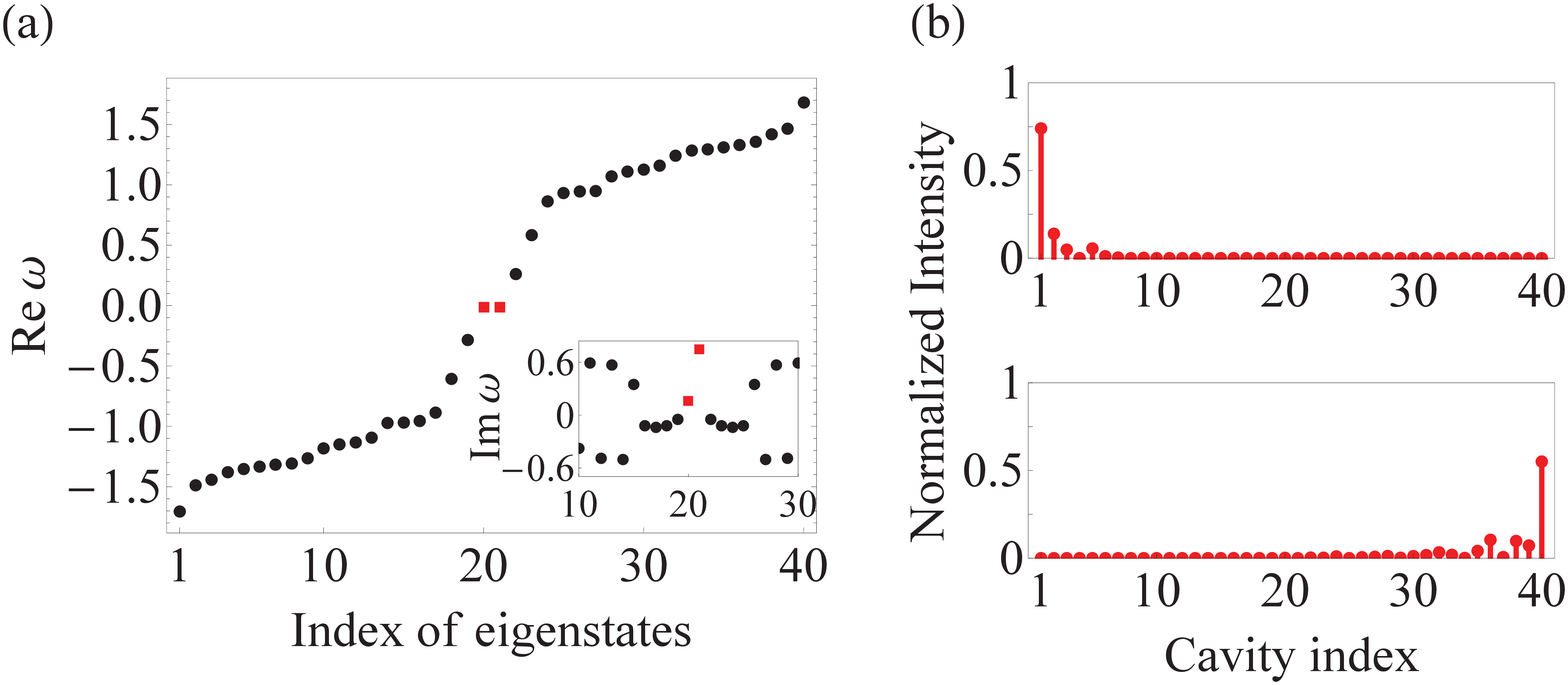}
\caption{Topological edge states under disorder in the cavity couplings, gain and loss. $g_1 = 1$, $g_2 = 0.5$ and $\sigma = 0.2$. (a) Sorted real parts of eigenvalues ${\rm Re} \, \omega$. Squares: edge stages stabilized at ${\rm Re} \, \omega = 0$. Inset: corresponding imaginary part ${\rm Im} \, \omega$. (b) Intensity distributions of the edge states.}
\label{fig:ESDO}
\end{figure}

Figure \ref{fig:ESDO} shows example eigenstates for a disordered lattice ($\sigma = 0.2$) in phase (II), based on the condition for Fig. 3 (b) in the main text.
The midgap states with ${\rm Re} \, \omega = 0$ are obtained even with a variation of 20\% in all the parameters [Fig. \ref{fig:ESDO} (a)].
Furthermore, the edge localization of these modes is clearly maintained [Fig. \ref{fig:ESDO} (b)].
This topological protection remains unless the bulk bandgap is effectively closed by the disorder.
In an experimental system with controlled gain and loss, the non-Hermitian effect for dimerization will be stably kept in practice.
Thus, variation in cavity couplings is the main factor in the symmetry preserving perturbation.
It is also noteworthy that possible cavity frequency detuning in each cavity will be a more serious obstacle, because it breaks the symmetries and hence lifts the edge-state frequencies toward the bulk band edges \cite{PhSSH}.

\section{Edge states in gapless phases}
The computation of the global Berry phase and its geometrical interpretation have clarified that the system in the gapless phases is also topologically nontrivial and trivial when $g_1 g_2 > 0$ $(W = 1)$ and $g_1 g_2 < 0$ $(W = 0)$, respectively.
Here, we define and discuss the non-Hermitian chirality for the states with ${\rm Re} \, \omega = 0$.
We then show the topological edge states in the gapless conditions and show what will become of these states for $W = 0$.

In terms of the pseudo-anti-Hermiticity in finite systems, $\hat{\mathcal{H}} = - \hat{\eta}' \hat{\mathcal{H}}^{\dagger} \hat{\eta}'$, a pair of dual left and right eigenstates based on a single non-Hermitian edge mode $\left| \psi \right\rangle$ with ${\rm Re} \, \omega = 0$ is related as $\hat{\eta}' \left| \psi \right\rangle = \pm \left| \psi \right\rangle \rangle$ \cite{NHedge}, where $\hat{\eta}' = \hat{\eta}'^{-1} = {\rm diag}(1,-1,1,-1, \dots, 1,-1,1,-1)$ in our system.
Here, including the normalization factor of the biorthonormal basis, we can define the edge-mode \textit{chirality} as,
\begin{equation}
\chi \equiv {\rm sgn} \left(\left\langle \psi \right| \hat{\eta}' \left| \psi \right\rangle \right) = \pm 1. \label{eq:chirality}
\end{equation}
$\chi$ physically distinguishes the intensity localization to the first and third cavities and that to the second and fourth cavities in all the unit cells.

Normally, the pseudo-anti-Hermiticity links two different bulk right eigenstates $\left| \phi \right\rangle$ and $\left| \phi' \right\rangle$ with detuning $\omega$ and $- \omega ^{*}$ by the duality.
However, edge states with ${\rm Re} \, \omega = 0$ generally have different ${\rm Im} \, \omega$ and cannot be mapped to any spatially detached states via the local operator $\hat{\eta}'$.
Thus, $\hat{\eta}'$ and the duality can only map $\left| \psi \right\rangle$ to itself.
In this sense, the chiral edge eigenstates with ${\rm Re} \, \omega = 0$ are \textit{independent} of each other (defective) and hence acquire a sort of robustness by the non-Hermitian symmetry.
The independence is based on the fact that the left eigenstate $\left| \psi' \right\rangle \rangle = \pm \hat{\eta}' \left| \psi \right\rangle$ that has an eigenvalue of $-\omega$ and the originally dual left state paired for $\left| \psi \right\rangle$, namely $\left| \psi \right\rangle \rangle$ with $\omega^{*}$, can coincide ($\left| \psi' \right\rangle \rangle = \left| \psi \right\rangle \rangle$) when ${\rm Re} \, \omega = 0$.
In contrast, for ordinary pairwise states $\{\left| \phi \right\rangle$, $\left| \phi' \right\rangle \}$ with ${\rm Re} \, \omega \ne 0$ and hence $\left| \phi' \right\rangle \rangle \ne \left| \phi \right\rangle \rangle$, the biorthonormality guarantees no chirality: $\left\langle \phi \right| \hat{\eta}' \left| \phi \right\rangle = 0$.

\begin{figure}[htbp]
\includegraphics[width=8.6cm]{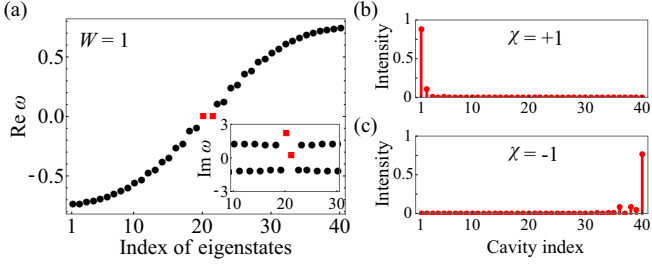}
\caption{Eigenstates of a topologically nontrivial forty-cavity system in gapless phase (IV). (a) Sorted eigendetuning ${\rm Re} \, \omega$. Inset: corresponding ${\rm Im} \, \omega$ profile. Squares: topological edge states with ${\rm Re} \, \omega = 0$. (b) Edge state localized at the left edge cavity with a chirality of $\chi = {\rm sgn} \left(\left\langle \psi \right| \hat{\eta}' \left| \psi \right\rangle \right) = + 1.$ (c) The other edge state confined at the right side with $\chi = -1$. $g_1 = 2.5$, $g_2 = 0.495$.}
\label{fig:edgeIVW1}
\end{figure}

Figure \ref{fig:edgeIVW1} shows eigenstates of a forty-cavity system in the gapless phase (IV), with $g_1 = 2.5$, $g_2 = 0.495$ and hence $W = 1$.
Here, $(g_1, g_2)$ is so close to the boundary between phase (III) and (IV) that only two discrete eigenstates can fall into ${\rm Re} \, \omega = 0$, as shown in Fig. \ref{fig:edgeIVW1} (a).
We then easily identify these two states as the topological edge states expected by $W = 1$, which strongly localize at the very left [Fig. \ref{fig:edgeIVW1} (b)] and right edge cavities [Fig. \ref{fig:edgeIVW1} (c)].
Their ${\rm Im} \, \omega$ values (net gain) are imbalanced due to the difference between $g_1$ and $g_2$ [Fig. \ref{fig:edgeIVW1} (a), inset].
They also have different chiralities $\chi$ [mode localization patterns, Eq. (\ref{eq:chirality})], meaning that they are independent.
We have made sure that such a pair of left and right edge states is obtained for the case of larger $g_1$, i.e. more eigenstates with ${\rm Re} \, \omega = 0$, and even in phase (V) giving flat real bands, as long as $W = 1$.

Nevertheless, we emphasize the recent observation \cite{1DTopoLaser2} that topological edge eigenmodes in gapless phases undergo significant mode mixing with bulk modes and lose both their spatial localization and spectral single mode features.
We hence expect that the midgap topological states in the main text are much more potential to be stabilized, especially in lasing operations.

\begin{figure}[htbp]
\includegraphics[width=8.6cm]{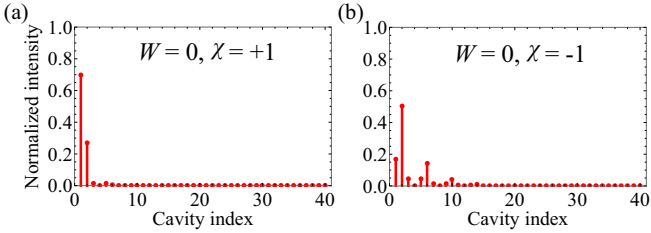}
\caption{Edge states in a system with $g_1 = 2.5$, $g_2 = -0.495$ [phase (IV)] and hence $W = 0$. (a) Edge state with $\chi = +1$ that continues to be on the left side. (b) Another edge state with $\chi = -1$, which has shifted from the right edge to the left. The global Berry phase $W$ denotes the difference of the number of edge states with different chiralities, $|n_{\chi +} - n_{\chi -}|$, on each side. The figures indicate the consistent result with the numerically computed $W$: $W = 1 - 1 = 0$ (left edge) and $W = 0 - 0 = 0$ (right edge).}
\label{fig:edgeIVW0}
\end{figure}

Next, we change $W$ from 1 to 0 by negating $g_2$.
The resultant system with $g_1 = 2.5$, $g_2 = -0.495$ also has two states with ${\rm Re} \, \omega = 0$, and their mode patterns are depicted in Fig. \ref{fig:edgeIVW0}.
Here, we find an edge state with $\chi = +1$ staying on the left edge [Fig. \ref{fig:edgeIVW0} (a)].
In contrast, another mode with $\chi = -1$ also localizes at the same side and distributes more into the bulk [Fig. \ref{fig:edgeIVW0} (b)].
Note that the same tendency holds for the case of phase (V).

The discussion in Ref. \onlinecite{NHedge} means that, $W$ determines the difference of the numbers of the edge modes with ${\rm Re} \, \omega= 0$ and different chiralities, $|n_{\chi +} - n_{\chi -}|$, at \textit{each} of the separate left and right edges (here, $n_{\chi +}$ and $n_{\chi -}$ denote the numbers of the edge states with $\chi = +1$ and $-1$).
$W = 1$ for Fig. \ref{fig:edgeIVW1} (b) and (c) hence indicates the single chiral left-edge and right-edge modes.
Fig. \ref{fig:edgeIVW0} with $W = 0$ then shows that no edge mode ($n_{\chi +} = n_{\chi -} = 0$) or a pair of edge states with opposite chiralities ($n_{\chi +} = n_{\chi -} = 1$) is allowed for every edge.
The imbalance in the numbers of the chiral edge states at the left and right sides is explained by nontrivial vortex charges of eigendetuning, in the next section.

\section{Vortex charges of eigenvalues}
Recently, fractional vortex charges of eigenvalues around exceptional points \cite{2DAnomoulousTI} have been found in a two-dimensional gapless system, and their potential contributions to edge states have been discussed.
For our one-dimensional system, we find a simple way to evaluate the vortex charge in $k$-space.
If the eigenvalue around an EP has the form with a complex phase factor: $\omega _{\rm EP} \equiv \omega (k \sim k_{\rm EP})  \propto \exp[i V {\rm Arg}({k - k_{\rm EP}})]$, we can consider a virtual loop in a complex plane of $k$ with an infinitesimal radius: $k = k_{\rm EP} + \delta r \exp(i \theta)$, where $\delta r \ll 1$ and $\theta$ is real.
We can then extract the vortex charge $V$ by performing a trivial integration along with $\theta$,
\\
\begin{align}
V =& \, \frac{1}{2 \pi} \lim_{\delta r \to 0} \oint \frac{\partial }{\partial k} \, {\rm Arg} [\omega (k_{\rm EP}, \delta r, \theta)] \, dk \nonumber \\
  =& \, \frac{1}{2 \pi} \lim_{\delta r \to 0} \int _{0}^{2 \pi} \frac{\partial }{\partial \theta} \, {\rm Arg} [\omega (k_{\rm EP}, \delta r, \theta)] \, d \theta \nonumber \\
  =& \, \frac{1}{2 \pi} \lim_{\delta r \to 0} \big\{ {\rm Arg} [\omega (k_{\rm EP}, \delta r, 2 \pi _{-})] \nonumber \\
   & \qquad \qquad \qquad \qquad - {\rm Arg} [\omega (k_{\rm EP}, \delta r, 0)] \big\},
\end{align}
where ${\rm Arg} [\omega (k_{\rm EP}, \delta r, 2 \pi _{-})]$ is the limit inferior for $\theta \rightarrow 2 \pi$.

The vortex charge is associated with the complex magnitude of the effective ``magnetic field" of the pseudospin system, which determines the eigendetuning \cite{2DAnomoulousTI}.
Here, the eigenvalues for the upper bands are re-denoted by,
\begin{equation}
\omega_{\pm} (k) = \frac{1}{\sqrt{2}} \sqrt{A \pm \sqrt{ A^2 - B^2 - 16 \kappa^4 \sin^2 \frac{k a}{2}}}, \label{eq:eigenvalue2}
\end{equation}
where the $\pm$ sign of $\omega_{\pm}$ stand for the corresponding sign of the right-hand side in Eq. (\ref{eq:eigenvalue2}). 
We then define $V_{\pm}$ as the vortex charge around $\omega_{\pm} (k_{\rm EP})$.
Because $\omega_{+}$ and $\omega_{-}$ are based on different \textit{magnitudes}, it is natural to define $V_{+}$ and $V_{-}$ as distinct indices for the vortices.
We notice that for the case of phase (III) and (V) [with a part of phase (I)], $k_{\rm EP}$ do not stay on the real axis but have finite imaginary parts.
Thus, $V_{\pm}$ in these conditions are obtained with such virtual values of $k_{\rm EP}$, out of the first Brillouin zone.
In addition, we find that $V$ is the same for $k = k_{\rm EP}$ and $-k_{\rm EP}$, and hence take $k_{\rm EP}$ here as the value with a positive real or imaginary part.
Note that the vortex charges for the lower bands can give artifactual values $\pm 1$ simply due to the discontinuity of ${\rm Arg}(\omega)$.
Except for the problem to be neglected, they have the same values for $V_{\pm}$, so we do not show them here.

\begin{figure}[htbp]
\includegraphics[width=8.6cm]{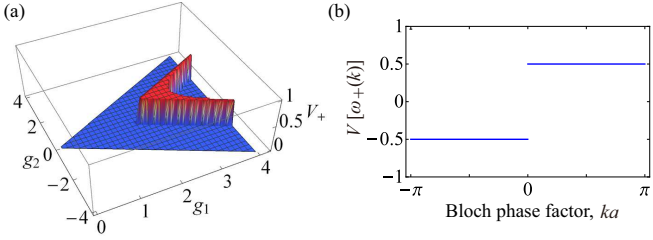}
\caption{(a) Vortex charge $V_{+}$ at $k = k_{\rm EP}$ for different $g_1$ and $g_2$. $V_{+} = 0$ for phase (I) ($g_1 < 2$, $g_1 > 2 \sqrt{2}$), (II), (III) and (V), $V_{+} = 1/2$ for phase (IV) [with phase (I), $2 < g_1 < 2 \sqrt{2}$]. $V_{+}$ is independent of the sign of $g_1 g_2$. $\delta r = 10^{-10}$. (b) Vortex charge $V[\omega_{+}(k)]$ of band eigenstates for phase (V). Despite the absence of EPs in the first Brillouin zone, the non-Hermitian flat band holds fractional vortex charges $\pm 1/2$, depending on the sign of $k$. $g_1 = 3.5$, $g_2 = 0.5$.}
\label{fig:VC}
\end{figure}

Figure \ref{fig:VC} (a) shows the dependence of $V_{+}$ on $g_1 > 0$ and $g_2$.
The vortex charge is constant in each band phase and is discretized as a multiple of 1/2, as shown previously \cite{2DAnomoulousTI}.
Importantly, $V_{+}$ is insensitive to the sign of $g_2$.
This indicates that $V$ is associated with a topological property distinct from that of the Berry phase $W$.

The vortex charges for the two EPs (based on $\omega_{\pm}$) are summarized in Table \ref{tab:VCEPS}.
Here, BZ means the inside the Brillouin zone, $0 \le k_{\rm EP} \le \pi$ for $k_{\rm EP} \ge 0$.
Finite fractional vortices are obtained only in phase (IV) [with phase (I), $2 < g_1 < 2 \sqrt{2}$], where we see the discontinuous jumps of eigenstates in the band structure [Fig. \ref{fig:Band3and4} (b)].

According to Ref. \onlinecite{2DAnomoulousTI}, $2V$ determines the number of non-Hermitian edge modes with ${\rm Re} \, \omega_{\rm EP} = 0$ and distinct chiralities.
Each of Fig. S7 [phase (IV), $W = 1$] and Fig. S8 [phase (IV), $W = 0$] shows a couple of edge states with different chiralities and ${\rm Im} \, \omega$.
This indicates that $V_{+} = 1/2$ and $V_{-} = -1/2$ are in charge of the single edge modes with $\chi = +1$ and $-1$, regardless of the value of $W$.
In contrast, the EPs are located in the separate upper and lower bands, ${\rm Re} \, \omega_{\rm EP} \ne 0$, in phase (I) ($g_1 < 2$), (II) and (III).
Thus, they cannot be relevant with the chiral edge states at ${\rm Re} \, \omega = 0$, resulting in $V = 0$.
We also see that $k_{\rm EP}$ in the extended complex plane are not likely to affect the system [$V = 0$, phase (III), (V) and (I) with $g_1 > 2 \sqrt{2}$].

Here, we find that $W$ and $V$ are complementary in our model.
While $W$ gives the difference in the number of the chiral edge modes on each edge, $V$ only describes their numbers in the entire system.
The finite $V_{+}$ and $V_{-}$ assure that the states in Fig. \ref{fig:edgeIVW0} with $\chi = +1$ and $-1$ under $W = 0$ do not annihilate each other, even though they are confined at a single side (\textit{anomalous} edge modes).
Such states are possible because the mirror symmetry is broken by non-Hermiticity.
\begin{table}[htb]
  \begin{center}
    \caption{Vortex charges of EPs ($g_1 > 0$)}
    \begin{tabular}{c|ccccc}
      phase & (I) ($g_1<2$) & (II)& (III) & (IV), part of (I)	  & (V), part of (I) \\ \hline 
      $k_{\rm EP}$ & BZ  & BZ & $i x$ & BZ & $\pi + i x$ \\ \hline
      $V_{+}$ & 0 & 0 & 0 & $\frac{1}{2}$ & 0 \\
      $V_{-}$ & 0 & 0 & 0 & $-\frac{1}{2}$ & 0 \\ 
    \end{tabular}
	\label{tab:VCEPS}
  \end{center}
\end{table}
\begin{figure}[htbp]
\includegraphics[width=8.6cm]{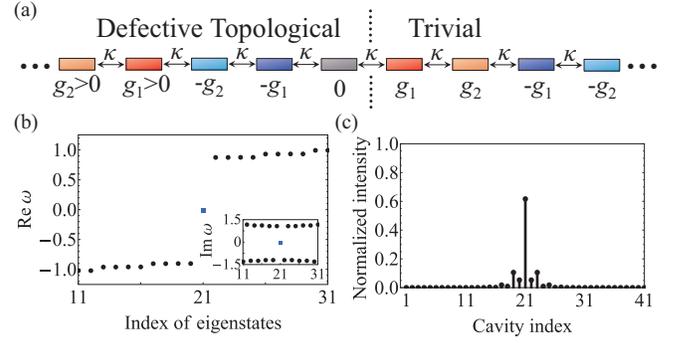}
\caption{(a) A system forming a topological bound state based on the global $\mathcal{PT}$ symmetry, which comprises a single vacant cavity and proposed trivial lattices with 20 cavities on both sides. (b) Selected ${\rm Re} \, \omega$ and ${\rm Im} \, \omega$ (inset). Square: the topological bound state with ${\rm Re} \, \omega = {\rm Im} \, \omega = 0$. (c) Intensity profile for the bound state. $g_1 = 2, g_2 = 1$.}
\label{fig:PTBS}
\end{figure}

Finally, what explains the non-Hermitian chiral edge states obtained in phase (V), where the system has a flat band in ${\rm Re} \, \omega (k) = 0$ and the vortex charge of the EP in the complex plane vanishes?
Interestingly, we find that a fractional vortex emerges not at the EP but in each bulk eigendetuning, for the case of phase (V) [Fig. \ref{fig:VC} (b)]. 
This is induced by the different ${\rm Im} \, \omega (k)$ of the bulk eigenstates for each $k$ in phase (V) [Fig. \ref{fig:Band3and4} (d)].

Vortex charges of EPs in the Brillouin zone, and those remaining in the entire non-Hermitian flat bands, would indicate chiral edge and defect states that are not necessarily based on the topological Berry charge \cite{TDSBulkTrivial1,TDSBulkTrivial2,TDSNH}.

\section{Topological bound state with global $\mathcal{PT}$ symmetry}
To achieve systematically a topological state with ${\rm Im} \, \omega = 0$, we place a cavity without gain or loss sandwiched between two 20-cavity lattices with different sequences of two-magnitude gain and loss, so that the whole lattice respects the global $\mathcal{PT}$ symmetry \cite{BoundPhTS} [Fig. \ref{fig:PTBS} (a)].
Here, the center vacant cavity and left parts can be considered a defective topological lattice, thus a bound state with ${\rm Re} \, \omega = {\rm Im} \, \omega = 0$ is formed [Fig. \ref{fig:PTBS} (b)].
Its mode profile shows localization at the vacant cavity [$n = 21$ in Fig. \ref{fig:PTBS} (c)].
Such a bound state is preserved under the combination of the pseudo-anti-Hermiticity and global $\mathcal{PT}$ symmetry, as shown in Ref. \onlinecite{BoundPhTS}.

As our final comments, we found that a similar structure to our system was included in an aperiodic lattice constructed according to Fibonacci Sequences \cite{DefectFibonacci}.
However, such a lattice basically includes large detuning in the real part of its refractive index profile for the design of its frequency response.
Thus, its concept is clearly different from our work.
Moreover, as described in Ref. \onlinecite{DefectFibonacci}, the defect states here are induced not by a nontrivial photonic topology but by the $\mathcal{PT}$ phase transition.
This can be seen by the fact that no edge state or defect state is reported in that system with the exact $\mathcal{PT}$ phase.


\begin{thebibliography}{87}
\bibitem [{Sup()}]{Supplementary}%
  \BibitemOpen
  \href@noop {} {}\bibinfo {howpublished} {See Appendices for describing the
  system gapless phases, bulk symmetry, effective decoupling, bandgap opening,
  topological number and its geometrical picture, possible finite size effects,
  robustness of the topological edge states, chiral edge states in gapless
  phases, vortex charges of eigendetuning with band coalescence, and
  non-Hermiticity-based topological bound state with $\mathcal{PT}$
  symmetry.}\BibitemShut {Stop}%
\end{thebibliography}
\end{document}